\documentclass[aps,prl,english,twocolumn,superscriptaddress]{revtex4-1}
\usepackage{graphicx}
\usepackage{amssymb}
\usepackage{amsmath, amstext, amssymb, amsfonts, amsxtra, mathtools, stmaryrd} 
\usepackage{braket}
\usepackage{textcomp}
\usepackage{relsize}

\usepackage[usenames,dvipsnames]{color}
\usepackage{units}
\usepackage[normalem]{ulem}

\newcommand{\su}{\uparrow} 
\newcommand{\sd}{\downarrow}

\newcommand{\bonn}{HISKP, University of Bonn, Nussallee 14-16, 53115 Bonn, Germany}
\newcommand{\singapore}{Science and Math Cluster and EPD Pillar, Singapore University of Technology and Design, 8 Somapah Road, 487372 Singapore}
\newcommand{\beheshti}{Department of Physics, Shahid Beheshti University, G.C., Evin, Tehran 19839, Iran}

\let\oldref\ref
\renewcommand{\ref}[1]{(\oldref{#1})}

\begin{document}
\title{Evolution of two-time correlations in dissipative quantum spin systems: aging and hierarchical dynamics}

\author{Stefan Wolff}
\affiliation{\bonn}
\author{Jean-S\'ebastien Bernier}
\affiliation{\bonn}
\author{Dario Poletti}
\affiliation{\singapore}
\author{Ameneh Sheikhan}
\email{Corresponding author: a\_sheikhan@sbu.ac.ir.}
\affiliation{\beheshti}
\affiliation{\bonn}
\author{Corinna Kollath}
\affiliation{\bonn}

\begin{abstract}
  We consider the evolution of two-time correlations in the quantum
  XXZ spin-chain in contact with an environment causing dephasing. Extending
  quasi-exact time-dependent matrix product state techniques to consider the dynamics of two-time correlations
  within dissipative systems, we uncover the full quantum behavior for these correlations
  along all spin directions. Together with insights from adiabatic elimination
  and kinetic Monte Carlo, we identify three dynamical regimes. For initial times, their evolution is dominated
  by the system unitary dynamics and depends on the initial state and the Hamiltonian parameters.
  For weak spin-spin interaction anisotropy, after this initial dynamical regime, two-time
  correlations enter an algebraic scaling regime signaling the breakdown of time-translation invariance
  and the emergence of aging. For stronger interaction anisotropy, these correlations first go
  through a stretched exponential regime before entering the algebraic one. Such complex relaxation arises due to
  the competition between the proliferation dynamics of energetically costly excitations
  and their motion. As a result, dissipative heating dynamics of
  spin systems can be used to probe the entire spectrum of the underlying Hamiltonian.  
\end{abstract}

\maketitle
%\date{\today}

% introduction
%
Two-time correlations are powerful tools to capture the fundamental dynamical features of many-body systems both in
and away from equilibrium. These correlation functions are of the form $\langle B(t_2) A(t_1) \rangle$
where $A$ and $B$ are operators, $t_1$ and $t_2$ are two different times,
and $\langle \dots \rangle = \text{tr}(\rho \dots)$ is the average over the
density matrix $\rho$ of a given system. 

Numerous experimental techniques have been developed to probe these correlations measuring the
response of many-body systems. A non-exhaustive list includes ARPES~\cite{LuShen2012},
neutron scattering~\cite{BramwellKeimer2014} or conductivity and magnetization
measurements in solids~\cite{AshcroftMermin}, and radio-frequency~\cite{StewartJin2008}, Raman,
Bragg~\cite{Esslinger2010} (and references therein) or modulation
spectroscopy~\cite{StoeferleEsslinger2004} in cold gases.
In equilibrium, these experimental methods provide information on various spectral features such as
collective excitations and bound states. Whereas
away from equilibrium, these techniques are employed to identify the formation of dynamically
induced states in isolated quantum systems subjected to an external parameter
change, and to capture, using for example magnetic susceptibility measurements, the aging dynamics of
classical spin glasses~\cite{VincentDupuis2017}.

Theoretically, two-time correlations have been studied in isolated many-body quantum systems
(i.e. not in contact with an environment), both in and far from equilibrium. However, for
open many-body quantum systems (i.e. in contact with an environment) evaluating out-of-equilibrium
two-time correlations has proven extremely challenging.  Most works have instead focused on characterizing
the non-equilibrium dynamics of open systems by considering the universal scaling behavior
of simpler observables or the propagation of
single-time correlations~\cite{MarinoSilva2012, BernierKollath2018}, by using various
approximate approaches to evaluate two-time
correlations~\cite{BuchholdDiehl2015, MarcuzziLesanovsky2015, SciollaKollath2015, EverestLevi2017, HeDiehl2017, DenisWimberger2018},
or by considering small many-body quantum systems~\cite{WangPoletti2018}.    

Here, for the first time, we evaluate quasi-exactly the 
evolution of both the two-time correlations along the $z$-spin
direction, $\langle S^{z}_{l}(t_2) S^{z}_{l+d}(t_1) \rangle$,
and along the $\pm$-spin directions, $\langle S^{+}_{l}(t_2) S^{-}_{l+d}(t_1) \rangle$,
in a quantum XXZ spin-$1/2$ chain
in contact with a memoryless environment causing dephasing.
$S^{z}_l$ and $S^{\pm}_l$ are the spin-$1/2$ operators in the $z$ and $\pm$
directions at site $l$. Previous works on this system had solely focused on
the evaluation of equal-time correlations along the $\pm$-spin directions~\cite{CaiBarthel2013}
identifying an algebraic regime similar to the one found for interacting
bosons in contact with a dissipative environment causing dephasing~\cite{PolettiKollath2012,PolettiKollath2013}.
Additionally, in the classical limit, for large interaction anisotropies, the
$\langle S^{z}_{l}(t) S^{z}_{l}(0) \rangle$ correlations were shown to display a stretched
exponential behavior~\cite{EverestLevi2017}.

We developed here a variant of the quasi-exact time-dependent variational matrix product
state (t-MPS) technique~\cite{White1992,Schollwoeck2011} applicable to two-time correlations.
Using this novel approach, we uncover the full quantum behavior of these correlations along both the
$z$ and the $\pm$-spin directions when the evolution of the spin system begins from an excited
state such as the N\'eel $|\!\!\su,\sd,\su,\sd,\dots,\su,\sd\rangle$
or the single domain wall state $|\su,\dots,\su,\su,\sd,\sd,\dots,\sd\rangle$.
Our analysis is carried out using an implementation of t-MPS built upon the ITensor library~\cite{itensor}.
The density matrix operator is represented as a pure state
in an enlarged Hilbert space and the evolution of
the two-time correlations is implemented using an approach similar to the one used
to obtain their equilibrium thermal counterparts~\cite{ZwolakVidal2004, VerstraeteCirac2009}.
Taking good quantum numbers into account~\cite{BonnesLaeuchli2014, BernierKollath2018}
enables us to follow the quasi-exact evolution of systems
for sufficiently long times to identify three interesting dynamical regimes signaled
by changes in the behavior of the two-time correlations along the $z$-spin direction.

The first regime, identified at initial times, is dominated by the system unitary dynamics.
This regime depends significantly on the initial state and on the Hamiltonian parameters, and,
for weak dissipation, resembles the dynamics of the isolated system.
The second regime, identified both numerically and analytically,
is characterized by an algebraic scaling: for distances $d \geq 1$, the
normalized correlations are proportional to $\left(t_2/t_1\right)^{-3/2}$ signaling
the emergence of aging dynamics with broken time-translation invariance.
In this particular case, this aging regime finds its origin in the presence of underlying
diffusive processes.
Finally, in the third regime, correlations evolve following a stretched exponential, a behavior typically associated
with glasses or systems exhibiting hierarchical separation of time-scales. This regime, which we identify for the
first time over a wide range of $t_1$ using quasi-exact simulations, only occurs for sufficiently
strong interaction anisotropies when the evolution begins from initial states with occupied energy levels well
separated from others. The evolution of these two-time correlations is governed by the competition between the
nucleation (or annihilation) dynamics of energetically costly excitations and
their motion, two processes occurring on very different time-scales.

We interpret our quasi-exact numerical findings using adiabatic elimination and kinetic Monte Carlo
from which the scaling properties can be predicted, and find the observed dynamics to closely relate to
the spectrum of the spin Hamiltonian. Monitoring the dynamics of two-time correlations
induced by dissipative heating thus constitutes
a novel approach to characterize spin systems as it reveals features spanning the entire spectrum of
the underlying Hamiltonian.
Let us also mention that the two-time correlations along the $\pm$-spin direction
decay exponentially and therefore exhibit a completely different behavior from which the regimes mentioned
earlier cannot be inferred.

% model
%
To investigate the non-equilibrium dynamics of two-time correlations in an open quantum system, we consider
a spin-$1/2$ chain under the effect of local dephasing noise. In such a
situation, the evolution of the density operator $\rho$ is described by the Lindblad master equation
\begin{align}
  \frac{\partial \rho}{\partial t} &= -\frac{i}{\hbar} \left[ H_{\text{XXZ}}, \rho \right] + \mathcal{D}(\rho).
  \label{eq:MasterEq}
\end{align}
The first term on the right-hand side describes the unitary evolution due to the XXZ spin-$1/2$ Hamiltonian
\begin{align}
H_{\text{XXZ}} = \sum_{j=1}^{L-1} \left[ J_{x} \left( S^x_j S^x_{j+1} + S^y_j S^y_{j+1} \right) + J_z S^z_j S^z_{j+1} \right],\nonumber
\end{align}
where $J_{x}$ and $J_{z}$ are the exchange couplings along the different spin directions, $S^\alpha_j$
is the $\alpha$-direction spin operator at site $j$, and $L$ is the length of the chain.
The isolated XXZ spin-chain is solvable by Bethe ansatz and
is known to present three distinct phases~\cite{MikeskaKolezhuk2004}: for $-1 \le J_z/J_x \le 1$, the easy plane anisotropic
phase is gapless, while $J_z/J_x <  -1$ presents a gapped ferromagnetic phase, and $J_z/J_x > 1$ hosts a gapped
antiferromagnetic phase. The second term on the right-hand side of Eq.~\ref{eq:MasterEq} describes the dephasing noise in Lindblad form
\begin{align}
\mathcal{D}(\rho) &= \gamma \sum_{j=1}^L \left(S^z_j \rho S^z_j  - \frac{1}{4} \rho \right),\nonumber
\end{align}
where $\gamma$ is the dissipation strength. This term acts like a source of heat inducing
spin fluctuations and eventually drives the system towards the infinite temperature state, the unique steady state
of the model. If not stated otherwise, we consider here a system initially prepared in the N\'eel state and investigate
its dynamics as the system is coupled to the environment and starts to undergo dephasing.
We access the full quantum dynamics by extending the quasi-exact t-MPS techniques available to dissipative
systems to the study of two-time correlations. To gain analytical insights
into the evolution of this system, we employ adiabatic elimination, valid for times larger than $1/\gamma$, 
to capture the dominant dissipative dynamics in the limit where $\hbar\gamma \gg J_z$.
We focus primarily on the two-time correlations
along the $z$-spin direction given by
\begin{eqnarray}
  \hbar^2~C_d(t_2,t_1) &=& \langle S^z_{\frac{L}{2}}(t_2) S^z_{\frac{L}{2}+d}(t_1) \rangle, \nonumber
\end{eqnarray}
where $d$ is the distance between two spins, as they provide the most insights into the dynamical properties
of the system, and we also investigate the corresponding equal-time correlations.

% figure: stretched exponential
\begin{figure}
\includegraphics[width=0.9\linewidth]{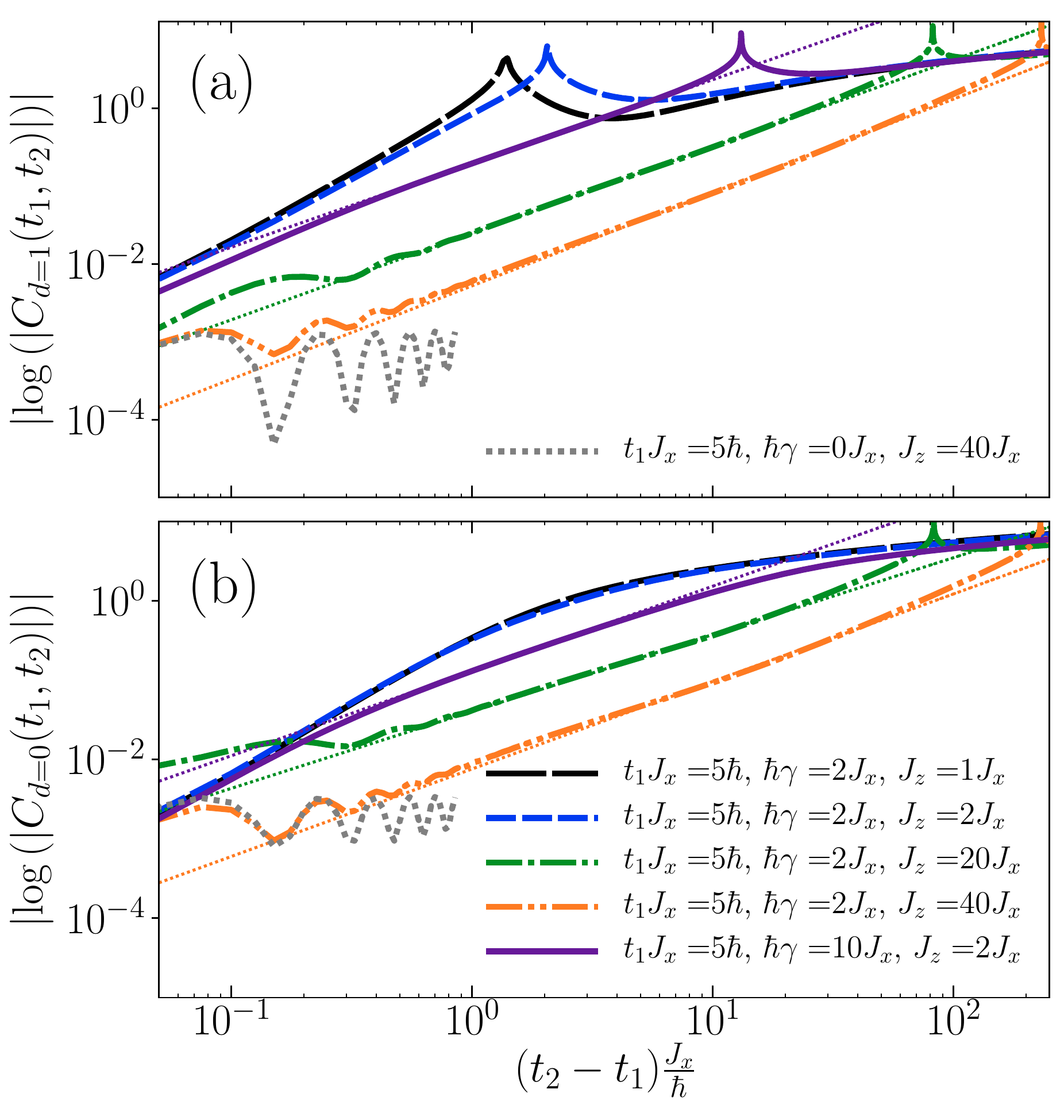}
\caption{\label{fig:stretched_exp} Initial quasi-unitary and intermediate stretched exponential regimes.
  Logarithm of the two-time correlation
  $\hbar^2 C_{d}(t_2,t_1) = \langle S^z_{\frac{L}{2}}(t_2) S^z_{\frac{L}{2}+d}(t_1) \rangle$
  for $d=1$, panel (a), and $d=0$, panel (b), versus the time difference $t_2-t_1$ (t-MPS data for $L=48$).
  For weak dissipation, the initial time-regime is dominated by the system unitary dynamics. The dephasing noise damps out
  the oscillations commonly present in the unitary evolution (dashed grey line). Following this initial regime, for sufficiently
  strong coupling anisotropy $J_z/J_x$, the two-time correlations follow a stretched exponential: the linear slopes
  indicate the presence of this regime. The thin dotted lines are fits in the linear regions.}
  %  t-MPS data shown for $L=48$}
\end{figure}

% figure: scaling collapse
\begin{figure}
\includegraphics[width=0.9\linewidth]{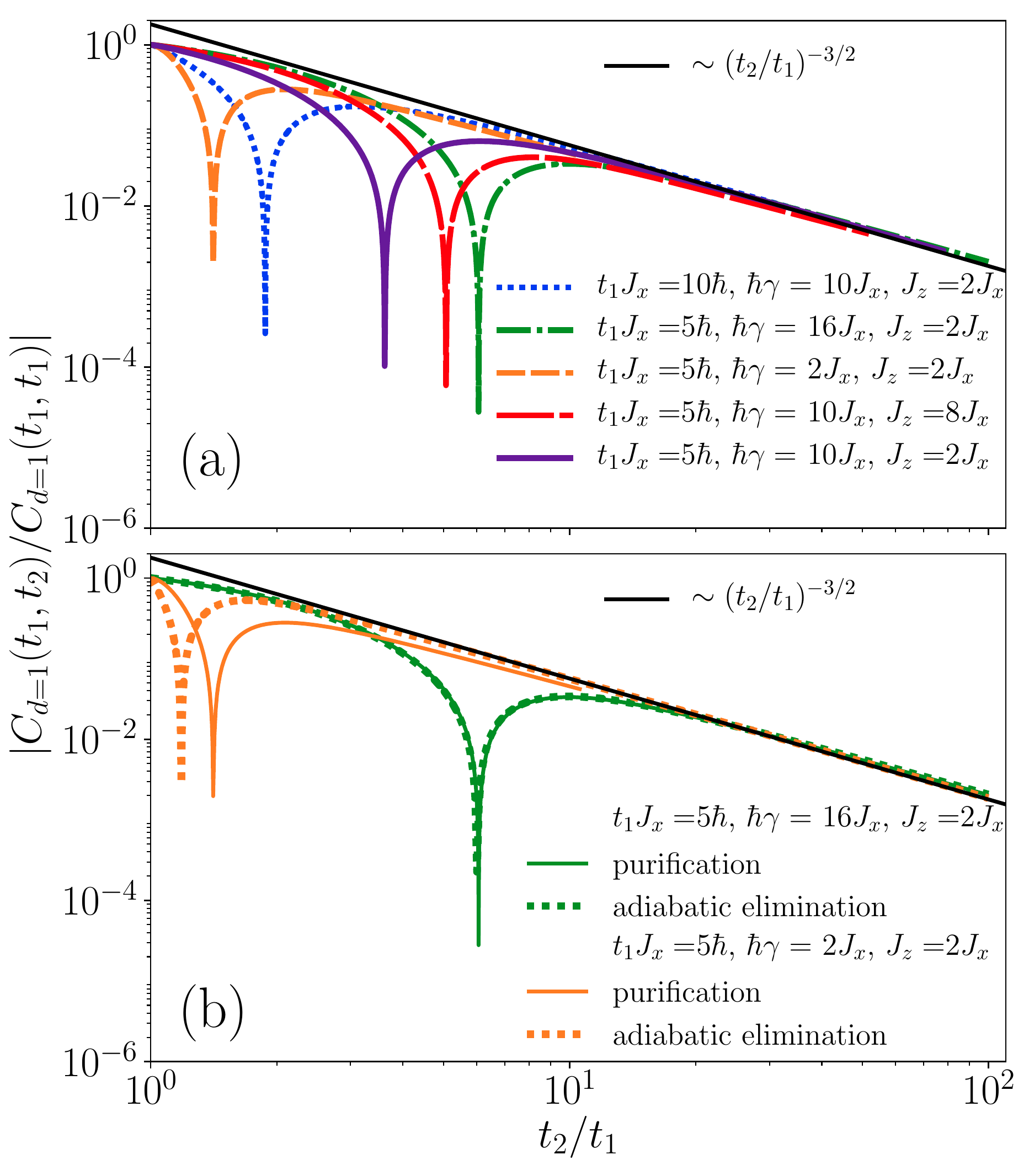}
\caption{\label{fig:aging} Scaling regime for the two-time correlations.
  Panel (a): Scaling collapse for the normalized two-time correlations between neighboring sites
  $\hbar^2 C_{d=1}(t_2,t_1) = \langle S^z_{\frac{L}{2}}(t_2) S^z_{\frac{L}{2}+1}(t_1) \rangle$
  calculated using t-MPS for $L=80$. The black solid line is a guide to the eye and highlights the
  $(t_2/t_1)^{-3/2}$ algebraic regime.
  Panel (b): Comparison between the two-time correlations obtained from t-MPS and
  adiabatic elimination. For large $\hbar\gamma/J_z$ ratios, adiabatic elimination describes 
  two-time correlations over the whole range of $t_2/t_1$ whereas, for smaller ratios, the analytical
  approach successfully captures the long-time $(t_2/t_1)^{-3/2}$ scaling but fails to describe
  the initial dynamics.}
\end{figure}

% discussion of results
%
As hinted earlier, we find the normalized two-time correlations along $z$-spin direction to present three distinct dynamical regimes
depending on the dissipation strength, $\hbar\gamma/J_x$, and the interaction anisotropy, $J_z/J_x$.
%
% unitary dynamics regime
%
For weak dissipation, the initial time-regime is governed by the system unitary dynamics. For strongly interacting systems,
the dynamics is typically characterized by oscillations due to the opening of a gap in the energy spectrum. Such dynamics, illustrated
by the grey dotted lines in Fig.~\ref{fig:stretched_exp}, is damped by the dephasing.
The other curves shown in Fig.~\ref{fig:stretched_exp} will be discussed in detail later. 

% scaling regime
%
After this initial unitary-like evolution, the system enters a scaling regime where the two-time correlations break time-translation
invariance as they do not depend on $t_2 - t_1$. This regime, which occurs at
later times for stronger interaction strengths, is exemplified in Fig.~\ref{fig:aging}
where the correlations for $d=1$ are shown. For $d=1$, the normalized two-time correlations scale as $\sim (t_2/t_1)^{-3/2}$ and one can see
that there is a regime for which curves with different $t_1$, $\gamma$, and $J_z$ nicely collapse on top of each other.
As this region is characterized by the slow algebraic relaxation of correlations, and by a dynamical scaling that is solely
a ratio of $t_2/t_1$, this system presents emergent aging dynamics.
As for $d=0$, we find in this case
that the two-time correlations also scale algebraically however they do not depend solely on a $t_2/t_1$ ratio.
As explained in the Supplemental material, these scaling regimes arise when $t_1$ lies within an interval where the
equal-time correlations along the $z$-spin direction decay algebraically.

To identify the origin of the scaling regime, we use many-body adiabatic
elimination~\cite{GarciaRipollCirac2009, PolettiKollath2012} to develop a set of differential equations
capturing the evolution around the dissipation-free subspace (see the Supplemental material for more details).
Then, resorting to the quantum regression theorem~\cite{GardinerZollerBook, BreuerPetruccione2002},
we find the two-time correlations along the $z$-spin direction to obey the following differential equations 
\begin{eqnarray}
  \frac{\partial}{\partial \tau} \langle S_j^z(t_1 + \tau) S_{j+d}^z(t_1) \rangle &=&
  \sum_l G_{j,l} \langle S_l^z(t_1 + \tau) S_{j+d}^z(t_1) \rangle \nonumber
\end{eqnarray}
where $\tau = t_2 - t_1$, $G_{j,l} = \frac{D}{2} (\delta_{j+1,l} + \delta_{j-1,l} - 2 \delta_{j,l})$, $D = \frac{J_x^2}{\hbar^2 \gamma}$
and the initial condition, $\langle S_j^z(t_1) S_{j+d}^z(t_1) \rangle$, is the solution of the equal-time correlations
at $t_1$~\cite{footnote1}. While these equations are in principle only valid for $J_z = 0$ and
more complicated expressions are obtained for finite $J_z$, we find that for sufficiently large $\hbar\gamma /J_z$
their solution and in particular the extracted scaling coincide well with the t-MPS results. As illustrated in Fig.~\ref{fig:aging} (b),
for large $\hbar\gamma /J_z$, adiabatic elimination describes two-time correlations
over the whole range of $t_2/t_1$, while for smaller ratios, this analytical approach describes well the long-time
$(t_2/t_1)^{-3/2}$ scaling but fails to capture the initial dynamics. The overall good agreement between
the t-MPS simulations and the adiabatic elimination approach within the scaling regime points to the diffusive
nature of the propagation of the two-time correlations under the action of dephasing. One should also note that this
regime is also present if the initial state is not the N\'eel state but is instead made of larger domains with alternating
magnetization.

Finally, the aging dynamics displayed by the correlations in the $z$-spin direction should be contrasted with the
evolution of the two-time correlations along the other spin directions.
For the latter, the evolution leaves the dissipation-free subspace through the application of the
lowering/rising operator $S^{\pm}_{l+d}$ at $t_1$. As a consequence, the dissipator strongly alters the evolution
and these correlations decay exponentially as a
function of $t_2 - t_1$: $\langle S_i^+(t_2) S_j^-(t_1) \rangle \propto e^{-\beta(\gamma) (t_2-t_1)}$
where $\beta$ is a function of the dissipative strength $\gamma$ (see Ref.~\cite{SciollaKollath2015}). 

% stretched exponential regime
%
Another interesting regime occurs solely at larger values of the interaction anisotropy $J_z/J_x$
and for particular initial states.
As shown in Fig.~\ref{fig:stretched_exp}, for intermediate values of the time difference, $t_2 - t_1$,
we find the two-time correlations along the $z$-direction
to follow a stretched exponential: $\log|C_{d}(t_2, t_1)| \sim (t_2-t_1)^{\nu_{d}}$
where $\nu_d$ depends on the system parameters.
We checked that this regime persists at least for distances up to $d = 9$ in a system of size $L = 48$.
This regime originates via the occurrence of nucleation events of energetically costly excitations.
For the disordered XXZ model, using classical approximations, a similar regime displaying a stretched exponential
decay was previously identified for the special case of $t_1=0$ where the two-time correlation reduces
to the single time staggered magnetization~\cite{EverestLevi2017}.
In comparison, here we identify, this regime for actual two-time correlation functions
over a wide range of $t_1$ using, for the first time, quasi-exact simulations within the t-MPS formalism
(see Fig.~\ref{fig:stretched_exp}).
Interestingly, even for large interaction anisotropy,
this stretched exponential regime gives way to the scaling regime discussed above at larger $t_2/t_1$ ratios.
These two contiguous regimes are displayed in Figs.~\ref{fig:stretched_exp} and \ref{fig:aging}
for the parameters $t_1 J_x = 5\hbar$, $\hbar\gamma = 10 J_x$ and $J_z = 2 J_x$.

\begin{figure}
\includegraphics[width=0.9\linewidth]{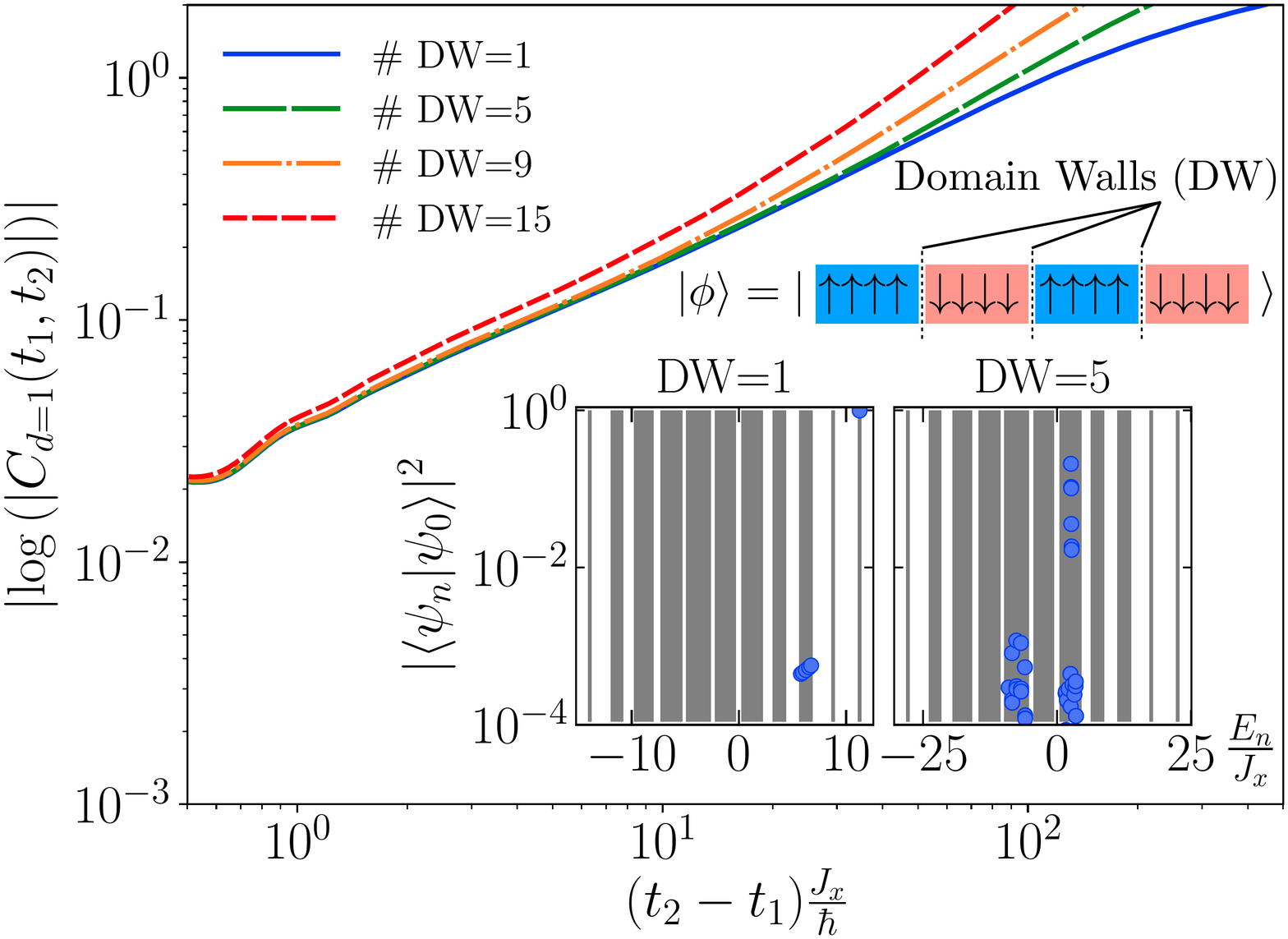}
\caption{\label{fig:domain_walls} Hierarchical dynamics. By increasing the number
  of domain walls (DW, see sketch in the figure) in the initial state, the dynamics of
  the two-time correlation function evolves faster from a regime where it follows a stretched exponential towards an algebraic region.
  The full quantum evolution obtained by t-MPS is shown for $L=64, J_z =10 J_x, \hbar \gamma = 2 J_x$ and $ t_1J_x =5 \hbar$.
  The inset shows the overlap (blue circles) of initial states with different numbers of domain walls with the energy
  eigenstates (grey vertical lines) of $H_{\text{XXZ}}$ computed with exact diagonalization for a system of size $L=12$ (with
  open boundary conditions) and $J_z =10 J_x$.}
\end{figure}

% crossover mechanism
%
The mechanism behind the crossover between the stretched exponential and the algebraic regimes can be
inferred by considering the proliferation of excitations caused by the dephasing noise.
We expect the stretched exponential regime to be dominant only if well separated time-scales exist for the nucleation
(or annihilation) of an excitation and for its motion. This separation typically occurs only for states
on the lower and upper bounds of the spectrum of the XXZ model (see the well separated energy bands at
the boundaries of the spectrum in the inset of Fig.~\ref{fig:domain_walls}). 

As the dissipative evolution brings the system into the infinite-temperature state, where all Hamiltonian
levels are equally occupied, we expect the stretched exponential to only show up in the initial dynamics when the
states at the boundaries of the spectrum are predominantly occupied. 
This situation explains the presence of the crossover from the stretched exponential to the algebraic regime seen
in Figs.~\ref{fig:stretched_exp} and \ref{fig:aging} ((a) panels) for the parameters
$t_1 J_x = 5\hbar$, $\hbar\gamma = 10 J_x$ and $J_z = 2 J_x$
where the initial state is the classical N\'eel state which, for $J_z > 0$, lies near the lower edge of the energy spectrum.
The time interval over which this regime occurs increases in width with the anisotropy
strength, since for large interaction anisotropies,
the difference between the intra and inter-band rates grows towards the edge of the spectrum. 
If the initial state is the N\'eel state, for $d=1$, we observe that its region
of existence terminates approximately when two-time correlation becomes zero for the first time.

To test this further, we consider different initial
states with zero total magnetization and a well defined number of domain walls. For the XXZ spin-chain with $J_z > 0$,
we first consider the state with one domain wall which should have the largest energy among this subset of states.
As illustrated in the inset of Fig.~\ref{fig:domain_walls}, for a system of $L = 12$ sites with a large
interaction anisotropy one finds, using exact diagonalization, that the state with one domain wall has indeed strong
overlap only with the most excited levels of the system and that these levels are all well separated from others by energy gaps.
The dephasing dynamics will then de-excite this state, but the rate of de-excitation to lower bands will be small
compared to the rate to change this state within its own band.
In contrast, considering the state with five domain walls, we find that it overlaps with levels located near
the center of the Hamiltonian spectrum,
where the energy bands are not well separated (see inset of Fig.~\ref{fig:domain_walls}).
In fact, for larger system sizes, these bands will get closer and closer together. 
In this case, there is no pronounced separation of scale between the intra and inter-band rate and the stretched exponential
regime will be absent. 

Comparing evolutions originating from states with increasing number of domain walls Fig.~\ref{fig:domain_walls}, we find that the
two-time correlations enter the stretched exponential regime only when the dissipative evolution begins from a state
on the outer edge of the spectrum
that is well separated in energy from other states (thanks to strong interactions) confirming the picture
detailed earlier. For initial states located within the center of the spectrum, where no clear separation of energy scales between
the band gaps and band widths is present, the evolution quickly enters the algebraic regime. Thus, using the initial state as a
knob, one can tune the system dissipative dynamics and unveil features of the entire underlying Hamiltonian spectrum.

% figure: generation/degeneration mechanism

% conclusion
%
In summary, considering the evolution of two-time correlations, we highlighted the extremely rich and intricate physics
at play in strongly interacting systems in contact with an environment. We evaluated quasi-exactly for the first time
these correlations along all spin directions extending dissipative MPS to two-time correlations, and showed
that their evolution is non-trivially affected by the presence of
a dissipative coupling, even leading to the breakdown of time-translation invariance. Perhaps most importantly,
we demonstrated that the dissipative heating dynamics reveals fundamental spectral features of the underlying Hamiltonian.
This finding paves the way to the development of non-equilibrium techniques to probe the spectrum
of strongly correlated many-body systems.

% acknowledgments
{\it Acknowledgments}: We thank I. Lesanovsky and M. Fleischhauer for enlighting discussions. We acknowledge
support from Shahid Beheshti University, G.C. (A.S.), Singapore Ministry of Education (D.P.),
Singapore Academic Research Fund Tier-II (project MOE2016-T2-1-065, WBS R-144-000-350-112) (D.P.)
and DFG (TR 185 project B4, SFB 1238 project C05, and Einzelantrag) and the ERC (Grant Number 648166) (C.K.).

\bibliography{swolff_twotime_arxiv}

\clearpage

%%%%%%%%%%%%%%%%%%%%%%%%%%%%%%%%%%%%%%%%%%%%%%%%%%%%%%%%%%%%%%%%%%%%%%%%%%%%%%%%%%%%%%%%%%%%%
% SUPPLEMENTARY

\setcounter{equation}{0}
\setcounter{page}{1}
\setcounter{figure}{0}

\onecolumngrid
\begin{center}
\Large
%%%%%%%%%%%%%%%%%%%%%%%%%%%
Supplemental material
%%%%%%%%%%%%%%%%%%%%%%%%%%%
\end{center}
\vspace{1cm}
\twocolumngrid

\subsection{Adiabatic elimination formalism}
\label{app:adiabatic_elimination}
At sufficiently large times, irrespective of the spin-spin interaction strength, the dissipation-free
subspace will be reached. While this subspace is highly degenerate with respect to the dissipator, the
Hamiltonian can possibly lift this degeneracy. In order to understand the non-equilibrium dynamics taking
place, we perform adiabatic elimination revealing how spin-flip induced virtual excitations around the
dissipation-free subspace affect the evolution of the system. For the system under study, in the presence
of periodic boundary conditions, the
dissipation-free subspace can be written down as
$\rho_0 = \sum_{\vec{\sigma}} \rho_{0,\vec{\sigma}} \vert \vec{\sigma}\rangle \langle \vec{\sigma}\vert$
where the different spin configurations are labeled within the $z$-component basis such
that $\vec{\sigma} = (\sigma_1, \sigma_2, \cdots, \sigma_L)$ with $\sigma_l = \pm 1/2$.
For times larger than $1/\gamma$, the density
matrix evolution is then effectively described by the set of differential equations
\begin{align}
  &\frac{\partial \rho_{0,\vec{\sigma}}}{\partial t} =
  \sum_{j=1}^{L}\frac{J_x^2 \gamma}{2\left[\left(J_z \alpha_j \right)^2 + \left(\hbar\gamma\right)^2\right]} \delta_{\sigma_j, \bar{\sigma}_{j+1}}
  \left( \rho_{0, \vec{\sigma}_j} - \rho_{0, \vec{\sigma}}\right), \nonumber
\end{align}
where $\alpha_j = 2 (\sigma_{j-1} \sigma_{j} +  \sigma_{j+1}  \sigma_{j+2})$, $\vec{\sigma}_j$ is
the spin configuration $\vec{\sigma}$ with swapped spins at site $j$ and $j+1$ and $\bar{\sigma}_j = -\sigma_j$. 

\subsection{Equal-time correlations}
\begin{figure}
\includegraphics[width=0.9\linewidth]{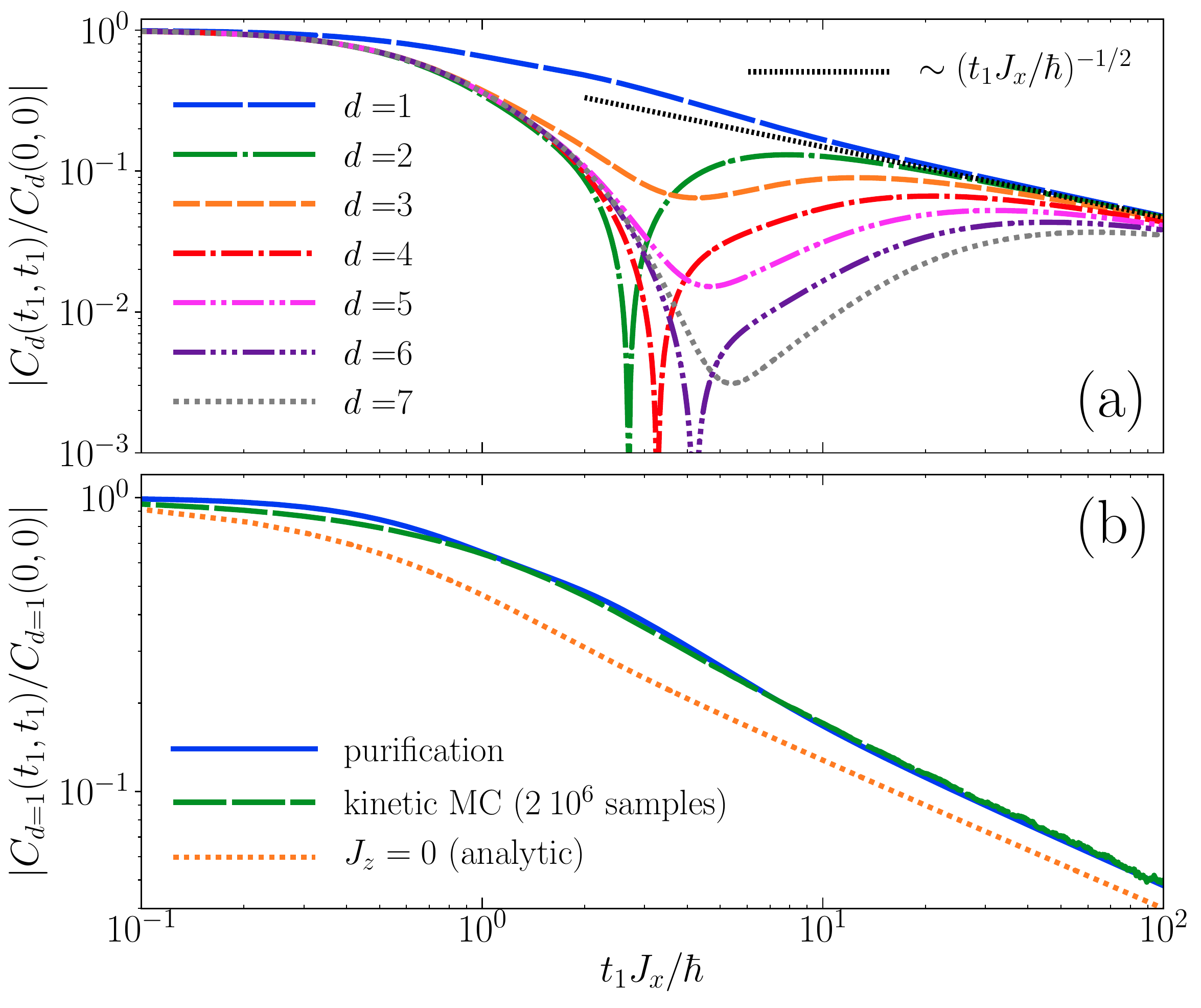}
\caption{\label{fig:diffusive} Equal-time correlations. Panel (a) shows the approach of a regime with diffusive dynamics for the
  normalized equal-time correlations, $|C_d(t_1,t_1)/C_d(0, 0)|$, for spins separated by different distances $d$ where
  $\hbar^2 C_d(t_1, t_1) = \langle S^z_{\frac{L}{2}}(t_1) S^z_{\frac{L}{2}+d}(t_1) \rangle$. The thin dotted black line is a guide to the eye
  highlighting the scaling at long times. Panel (b) compares the full quantum evolution of the density matrix, with a
  kinetic Monte Carlo simulation of the evolution within the dissipation-free subspace, and with the solution obtained by
  solving the differential equations presented at Eq.~\ref{eq:diffusion1t}. Parameters are $\hbar \gamma = 2 J_x, L=80$ and $J_z = 2 J_x$.}
\end{figure}

Within adiabatic elimination, the
equal-time correlations can be calculated in two different ways. Using
kinetic Monte Carlo, we can solve numerically for $\rho_0$ and then compute the correlations. While in
a second approach, valid for $\hbar\gamma \gg J_z$, we use the differential equation found above for $\rho_0$
to write down a set of coupled differential equations for $\hbar^2 C_{j,j+d}(t_1, t_1) = \langle S^z_{j}(t_1) S^z_{j+d}(t_1) \rangle$.
Together with periodic boundary conditions, these equations take the form
\begin{align}
  \frac{\partial}{\partial t_1}C_{j,j \pm 1}(t_1, t_1) &= \frac{D}{2} \left(C_{j \mp 1,j \pm 1} + C_{j,j \pm 2} - 2 C_{j,j \pm 1} \right), \nonumber \\
  \frac{\partial}{\partial t_1}C_{j,j+d}(t_1, t_1) &= \frac{D}{2} \left(C_{j+1,j+d} + C_{j-1,j+d} + C_{j,j+d+1} \right. \nonumber \\
  & \qquad \left. +~C_{j,j+d-1} - 4 C_{j,j+d} \right),~~~~\text{for $|d|>1$}, \nonumber \\
  \label{eq:diffusion1t}
\end{align}
where $D = \frac{J_x^2}{\hbar^2\gamma}$ and here $C_{l,l+d}$ stands for $C_{l,l+d}(t_1,t_1)$.
If the system is initially prepared in the N\'eel state, the correlations are translationally invariant,
$C_{d}(t_1,t_1) = C_{j,j+d}(t_1,t_1)$ with equations
\begin{align}
  \frac{\partial}{\partial t_1}C_{\pm 1}(t_1, t_1) &= D \left(C_{\pm 2} - C_{\pm 1} \right), \nonumber \\
  \frac{\partial}{\partial t_1}C_{d}(t_1, t_1) &= D \left(C_{d+1} + C_{d-1} - 2 C_{d} \right),~~~~\text{for $|d|>1$}, \nonumber 
\end{align}
and one should note that $C_{d} = C_{-d}$. To solve this system of differential equations, it is advantageous
to redefine the equal-time correlations such that the evolution for all distances is decribed by an
differential equation of the same form. To do so, we redefine the correlations as
$\tilde{C}_d(t_1, t_1) = C_d(t_1, t_1)$ for $d \ge 1$ and $\tilde{C}_{d+1}(t_1, t_1) = C_d(t_1, t_1)$ for $d\le -1$ implying that
$\tilde{C}_d(t_1, t_1) = \tilde{C}_{-d+1}(t_1, t_1)$ for $d\ge 1$. One
can then write down a diffusion equation for $\tilde{C}_{d}$ with diffusion constant $D$ and periodic boundary condition
\begin{align}
  \frac{\partial}{\partial t_1}\tilde{C}_{d}(t_1, t_1) =  D \left(\tilde{C}_{d+1} + \tilde{C}_{d-1} - 2 \tilde{C}_{d} \right) \nonumber
\end{align}
valid for $-\frac{L}{2}+2\leq d\leq \frac{L}{2}$.
This equation can be solved analytically in terms of the modified Bessel functions $I_n(x)$, and has for solution
\begin{align}
  & \tilde{C}_{d}(t_1, t_1) = \frac{1}{4}e^{-2Dt_1}~~\times \nonumber \\
  &\quad \left(-I_{d}(2Dt_1) + \sum_{j=-\frac{L}{2}+2}^{\frac{L}{2}} (-1)^j \text{sign}(j) I_{d-j}(2Dt_1) \right). \nonumber
\end{align}
For $d\ge 1$, in the limit where $L \gg 1$, the equal-time correlations take the form
\begin{align}
  \label{eq:Ct1t1} 
  & C_{d}(t_1, t_1) = \frac{(-1)^d}{4} e^{-2Dt_1} \sum_{j=1-d}^{d-1} (-1)^j I_{j}(2Dt_1),
\end{align}
and, furthermore, in the long-time limit, $Dt_1 \gg 1$, when $I_n(x) \sim e^x/\sqrt{2\pi x}$,
these correlations simplify to
\begin{align}
  & C_{d}(t_1, t_1) \sim -\frac{1}{\sqrt{64\pi D t_1}}. \nonumber
\end{align}
Therefore,  equal-time correlations scale in time as $t_1^{-1/2}$ in agreement with t-MPS
simulations as seen in Fig.~\ref{fig:diffusive} (a).
In fact, due to the form of the differential equations, one can infer that equal-time correlations propagate diffusively under
the action of the dephasing environment. In Fig.~\ref{fig:diffusive} (b), one sees that all three methods, kinetic Monte Carlo,
analytical adiabatic elimination and t-MPS, predict the same scaling behavior at large times. While parallel,
the analytical curve appears slightly below the two other ones, this discrepancy arises as obtaining an analytical solution
requires to set $J_z = 0$. However, even for finite $J_z$, the agreement between the
analytical and the two other solutions gets better as the ratio $\hbar \gamma/J_z$ increases.

% equal-time correlations

\subsection{Two-time correlations}
The long-time scaling of two-time correlations can also be understood analytically. In this case, to make progress, one needs to
resort to adiabatic elimination and also make use of the quantum regression theorem~\cite{GardinerZollerBookSupp, BreuerPetruccione2002Supp}.
As within adiabatic elimination, for times larger than $1/\gamma$ and for $\hbar\gamma \gg J_z$, the evolution of $\langle S^z_{j}(t) \rangle$
is governed by the linear differential
equation
\begin{align}
  \frac{\partial}{\partial t_1} \langle S^z_{j} \rangle &= \nonumber
  \frac{D}{2} \sum_{l=1}^{L} \left(\delta_{j+1,l} + \delta_{j-1,l} - 2 \delta_{j,l} \right) \langle S^z_{l} \rangle \nonumber \\
  &= \sum_{l=1}^{L} G_{j,l}~\langle S^z_{l} \rangle, \label{eq:Sz}
\end{align}
the quantum regression theorem states that the two-time correlation functions
\begin{align}
  \hbar^2 C_{j,j+d}(t_1 + \tau, t_1) = \langle S^z_{j}(t_1 + \tau) S^z_{j+d}(t_1) \rangle \nonumber
\end{align}
should be described by the differential equations
\begin{eqnarray}
\frac{\partial}{\partial \tau} C_{j,j+d}(t_1 + \tau, t_1) &=& \sum_{l=1}^{L} G_{j,l}~C_{l,j+d}(t_1 + \tau, t_1) \nonumber
\end{eqnarray}
where $G_{j,l}$ are the same matrix elements as in Eq.~\ref{eq:Sz}. Assuming once again spatial translation
invariance this set of equations reduces to a smaller set of diffusive equations for $C_d(t_1 + \tau, t_1)$ with diffusion constant $\frac{D}{2}$,
\begin{align}
\frac{\partial}{\partial \tau} C_{d}(t_1 + \tau, t_1) =\frac{D}{2}(C_{d+1}+C_{d-1}-2C_{d}), \nonumber
\end{align}
here $C_{l}$ stands for $C_{l}(t_1+\tau,t_1)$. Solving this set of differential equations, 
we find the two-time correlations along the $z$-direction to evolve as
\begin{align}
  C_{d} (t_2,t_1) &= e^{-D (t_2-t_1)}~~\times \nonumber \\
  &\quad \quad \sum_{d'=-\frac{L}{2}+1}^{\frac{L}{2}} C_{d'}(t_1, t_1)~I_{d-d'}(D(t_2 - t_1)) \nonumber
\end{align}
where $t_2=t_1+\tau$. Then, using as initial conditions $C_{d}(t_1, t_1)$ obtained in Eq.~\ref{eq:Ct1t1}
for $d \ge 1$ together with $C_{d=0}(t_1, t_1) = \frac{1}{4}$ and $C_{-d}(t_1, t_1)=C_{d}(t_1, t_1)$, we find
for $d \ge 0$, in the limit $L \gg 1$, that the two-time correlations can be rewritten in the
more amenable form
\begin{align}
  &C_{d}(t_2,t_1) = \frac{1}{4} e^{-D (t_2-t_1)} I_d(D(t_2-t_1)) \label{eq:solCdt2t1} \\
  &-\frac{1}{4}~\delta_{0,d}~e^{-D (t_2+t_1)} I_0(D(t_2+t_1)) \nonumber \\
  &+\frac{(-1)^d}{4} (1-\delta_{0,d})~e^{-D (t_2 + t_1)} \sum_{j=1-d}^{d-1} (-1)^j I_j(D(t_2+t_1)) \nonumber \\
  &+G_d(t_2,t_1), \nonumber
\end{align}
where
\begin{align}
  \label{eq:Gd}
  G_d(t_2,t_1) &= e^{-D (t_2-t_1)} \sum_{d'=1}^\infty C_{d'}(t_1,t_1) \\
  &\times~(I_{d+d'}(D(t_2-t_1)) - I_{d+d'-1}(D(t_2-t_1))). \nonumber
\end{align}
Using this expression, we then evaluate the scaling of the normalized two-time correlations in the limit
$D t_2 \gg D t_1 \gg 1$. While for the first three terms of Eq.~\ref{eq:solCdt2t1}, we simply expand
$I_n(x)$ for large $x$, for $G_d(t_1, t_2)$, we also need to take the continuum limit in order to carry out
analytically the sum over $d'$. This
additional limit amounts to approximate $C_{d'}(t_1,t_1)$ in Eq.~\ref{eq:Gd} as
\begin{align}
 &C_{d'}(t_1,t_1) \sim -\frac{1}{4} \frac{1}{\sqrt{2\pi (2 D t_1)}}~e^{-\frac{1}{2}\frac{d'^2}{2 D t_1}}. \nonumber
\end{align}
For $|d| \ge 1$, the normalized two-time correlations therefore scale as
\begin{align}\label{eq:2timeCorrlimit}
  \frac{C_{d}(t_2,t_1)}{C_{d}(t_1,t_1)} \sim & -\sqrt{2}~\left(\frac{t_2}{t_1}\right)^{-\frac{3}{2}} \times \nonumber \\
  & \left(1 + \frac{1}{\sqrt{\pi}} (D t_1)^{-\frac{1}{2}} - \frac{1}{4} (D t_1)^{-1} \right). \nonumber
\end{align}
Thus, for very large $D t_1$, only the leading contribution remains and the normalized two-time correlations scale as
\begin{equation}
\left|\frac{C_{d}(t_2,t_1)}{C_{d}(t_1,t_1)}\right| \sim \sqrt{2}\left(\frac{t_2}{t_1}\right)^{-\frac{3}{2}},\quad |d|\ge 1, \nonumber
\end{equation}
which is in agreement with the results obtained from t-MPS. This result highlights that aging dynamics can emerge from
diffusive processes triggered by dephasing noise.
Finally, for $d=0$, where $C_0(t_1, t_1)=\frac{1}{4}$, the long-time limit of the normalized two-time correlation scale as
\begin{align}
  \frac{C_0(t_2, t_1)}{C_0(t_1, t_1)} \sim & \frac{1}{\sqrt{2\pi}}~(D t_1)^{-\frac{1}{2}}\left(\frac{t_2}{t_1}\right)^{-\frac{3}{2}} \times \nonumber \\
  & \left( 1 + \frac{1}{\sqrt{\pi}} (D t_1)^{-\frac{1}{2}}
    - \frac{1}{4} (Dt_1)^{-1} \right). \nonumber
\end{align}
Consequently, on-site two-time correlations break time-translational invariance and scale algebraically; however, to leading order, these correlations
do not solely depend on the ratio $t_2/t_1$ and thus do not display aging.

\end{document}